\documentclass[aps, prl, showpacs, twocolumn, amsfonts, amsmath, amssymb, superscriptaddress, floatfix]{revtex4}

\usepackage[T1]{fontenc}
\usepackage[latin9]{inputenc}
\usepackage{color}
\usepackage{graphicx}

\def\be{\begin{equation}}
\def\ee{\end{equation}}
\def\bea{\begin{eqnarray}}
\def\eea{\end{eqnarray}}
\def\ra{\rangle}
\def\la{\langle}

\begin{document}

\title{Many-body Matter-wave Dark Soliton}

\author{Dominique Delande}
\affiliation{Laboratoire Kastler Brossel, UPMC-Paris6, ENS, CNRS; 4 Place Jussieu, F-75005 Paris, France}

\author{Krzysztof Sacha} 
\affiliation{
Instytut Fizyki imienia Mariana Smoluchowskiego, 
Uniwersytet Jagiello\'nski, ul.~Reymonta 4, PL-30-059 Krak\'ow, Poland}
\affiliation{
Mark Kac Complex Systems Research Center, 
Uniwersytet Jagiello\'nski, ul.~Reymonta 4, PL-30-059 Krak\'ow, Poland}


\begin{abstract}
The Gross-Pitaevskii equation -- which describes interacting bosons in the mean-field approximation -- possesses solitonic solutions in dimension one. 
For repulsively interacting particles, the stationary soliton is dark, i.e. is represented by a local density minimum. 
Many-body effects may lead to filling of the dark soliton. Using quasi-exact many-body simulations, we show
that, in single realizations, the soliton appears totally dark although the single particle density tends to be uniform.
\end{abstract}

\maketitle

Non-linear wave equations can describe localized disturbances which propagate without change of form. They are named solitary waves, or solitons, and appear in many different systems. 
Solitons are studied in non-linear optics: a localized pulse (bright soliton \cite{Zakharov71}) or a dark hole created on the continuous-wave background (dark soliton \cite{Zakharov73}) can propagate without spreading in a self-focusing and self-defocusing Kerr medium, respectively \cite{KivsharOpticalSol}. 
Matter wave dark and bright solitons have been realized experimentally in ultra-cold atoms \cite{burger1999,denschlag2000,strecker2002,khaykovich2002,becker,yefsah2013}. 
Within the mean field approach and at zero temperature, an atomic Bose-Einstein condensate (BEC) is described by the Gross-Pitaevskii equation (GPE). 
In one-dimensional (1D) space, this non-linear wave equation possesses a dark (bright) soliton solution if particle interactions are effectively repulsive (attractive). 
The GPE assumes all particles in the same single particle state and neglects that the interactions can populate other modes. 

There is a debate in the literature concerning many-body effects in a dark soliton \cite{corney97,corney01,dziarmaga2002,law2003,dziarmaga2003,dziarmaga2006,
martin2010a,martin2010b,dziarmaga2004,mishmash2009a,
mishmash2009b,dziarmaga2010,mishmash2010,krutitsky2010,sato12a,sato12b}. 
The GPE predicts a stable dark soliton state in 1D. The Bogoliubov corrections show that signatures of the soliton tend to disappear 
in the single particle density because particles depleted from the condensate fill the soliton notch \cite{dziarmaga2002,law2003,dziarmaga2003,dziarmaga2006}. 
However, the single particle density refers to results averaged over many experimental realizations and is not able to predict a single experimental outcome. 
Within the Bogoliubov approach, it is possible to simulate single experiments: in a single photo, 
the soliton is completely dark but its position varies randomly from realization to realization \cite{dziarmaga2003,dziarmaga2006}. 
This is a signature of the quantum character of the soliton, i.e. the soliton position is not described by a classical variable but by a probability density.
The Bogoliubov analysis is limited to quantum fluctuations of the soliton position on a scale smaller than the healing length and cannot predict what happens when many-body effects become stronger. 
The latter situation has been considered in 
Refs.~\cite{mishmash2009a,mishmash2009b} within many-body numerical calculations. 
Analysis of the second order correlation function leads the authors to the conclusion that a dark soliton cannot be observed when the single particle density becomes uniform. 
However, full simulations of single experiments involve much higher order correlation functions \cite{dziarmaga2010,mishmash2010}.
It is the goal of the present work to perform many-body numerical simulations of the entire experiment starting from soliton preparation 
till the destructive measurement of all particle positions. 
We show that a fully dark soliton can be observed when many-body effects are strong, although its measured position is random and varies among
experimental realizations.
In the following we assume the zero temperature limit and do not consider thermodynamical instability of the dark soliton \cite{muryshev2002,karpiuk2012}.

The stationary solution of the 1D GPE for particles of mass $m$ and interaction coefficient $g_0,$ associated with a dark soliton localized at $x=0$ reads
\be
\phi_0(x)=\sqrt{\rho}\tanh \frac{x}{\xi},
\label{dsol}
\ee
where $\rho$ is the particle density and $\xi=\hbar/\sqrt{mg_0\rho}$ the healing length \cite{Zakharov73,Kivshar98}.
A soliton localized at position $q$ is
represented by $\phi_0(x-q).$
In this mean-field approach, the position $q$ of the soliton is just a parameter, i.e. a classical variable.
Beyond the mean-field approximation, deviations for a pure dark soliton must be taken into account and $q$ becomes a quantum degree of freedom.
In the Bogoliubov approach, one linearizes the many-body equations in the vicinity of the GPE solution. 

If the cloud of particles is trapped in a 1D harmonic potential of low frequency $\omega$, $\phi_0(x)$ is a good approximation for the stationary dark soliton solution of the GPE in the vicinity of the trap center. The Bogoliubov approach shows that, apart from usual positive energy quasi-particle modes, there is one anomalous mode with negative energy, corresponding
to the translation of the soliton position. This mode is strongly localized in the soliton notch because its dominant part is the translation mode of the soliton, $\phi_{tr}(x)\propto \partial_x\phi_0(x)$ \cite{dziarmaga2002}. Consequently, the single particle density, corresponding to the stationary dark soliton, is not zero at the soliton position because there are particles which occupy the anomalous mode. 
In the thermodynamic limit  $\omega\rightarrow 0$ but $N\omega=$const, the number 
of particles in the anomalous mode diverges like $1/\omega$ which indicates breakdown of the Bogoliubov approach \cite{dziarmaga2002}. This limit shows also that the stationary dark soliton state in the homogeneous case cannot be described by the Bogoliubov theory.

The dark soliton is not the ground state of the system and therefore cannot be obtained by cooling an atomic gas.
In order to prepare a dark soliton, a phase imprinting method is used in experiments \cite{burger1999,denschlag2000,becker,yefsah2013}. 
Starting with an atomic gas in the ground state, half of the cloud acquires a phase $\pi$ after a short interaction with a laser radiation. 
This phase difference leads quickly to the formation of a dark soliton. 
In the ideal situation all atoms are prepared in the same single particle state (\ref{dsol}). 
Such a perfect condensate is not an eigenstate of the system: during
the many-body time evolution, the localization of the soliton at its initial position
is progressively lost, with growing occupation of the translation mode and 
a progressive depletion of the condensate. 
The Bogoliubov approach may represent accurately the initial evolution but breaks down when the soliton delocalization becomes comparable to the healing length $\xi$.

\begin{figure}
\includegraphics[width=0.9\columnwidth]{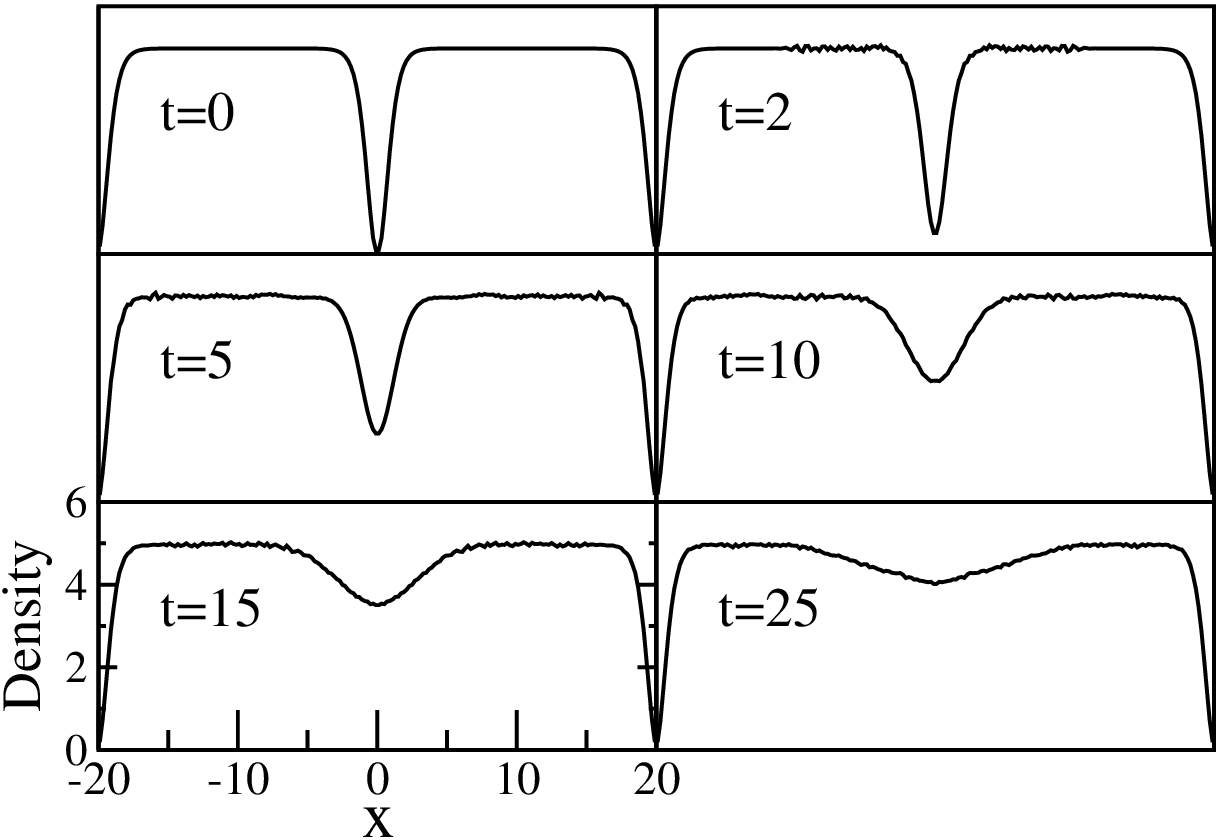}
\caption{(color online) Temporal evolution of a dark soliton in a box. The soliton is prepared by starting from the ground state in the presence of a barrier located in the middle of the box which creates a large density hole. At $t=0$ the barrier is turned off and a phase is imprinted. The subsequent dynamics is computed using a quasi-exact many-body TEBD algorithm. 
Different panels show the one-body density at increasing time.
At short time the soliton notch emits density waves (visible as small modulations
of the otherwise flat background) and is progressively filled at longer times. The density notch also spatially spreads producing an almost uniform density.
In this figure and the following ones, $x$ is in units of the healing length $\xi.$}
\label{spdensity}
\end{figure}

In order to describe the soliton delocalization, we perform quasi-exact many-body numerical simulations.
In the following, $\xi$ is chosen as the unit of length and the chemical potential as the energy unit.
We consider $N=180$ atoms in a box of length $L=40$ with open boundary conditions and $g_0=0.2,$
so that the background density is $\rho=5.$ The corresponding parameter for the Lieb-Liniger model~\cite{sato12a} is $\gamma=0.04,$ indicating the weakly interacting regime. 
Space is discretized on a grid with $\Delta x=0.2$, so that the many-body Hamiltonian in the second-quantization formalism is a Bose-Hubbard Hamiltonian with tunneling amplitude $1/(2\Delta x^2)$ and interaction energy $g_0/\Delta x$ \cite{Schmidt2007,mishmash2009a,mishmash2009b,delande2013} \footnote{The space discretization introduces a cut-off in particle momenta which for $\Delta x=0.2$ corresponds to $\Delta p = \pm \pi/\Delta x \approx \pm 15.7$ in the units used in this letter. $\Delta x$ is chosen sufficiently small for discretization effects to be extremely small, if not negligible, in our calculations}. 
An $N$-body state of the system can be represented by a matrix product state (MPS) 
 \begin{equation}
|\psi\rangle\! = \!\!\!\!\!\!\sum\limits_{\genfrac{}{}{0pt}{}{\alpha_1,\ldots,\alpha_M}{i_1,\ldots,i_M}}\!\!\!\!\!
\Gamma_{1\alpha_1}^{[1]
,i_1}\lambda^{[1]}_{\alpha_1}\Gamma^{[2],i_2}_{\alpha_1\alpha_2}\ldots\Gamma^{[M],i_M}_{\alpha_{n-1}1} |i_1,\dots,i_M\rangle,
\label{eq:MPS}
\end{equation}
where $|i_1,\dots,i_M\rangle$ is a Fock state with definite numbers of atoms in each of $M$ sites of the discrete space, $\Gamma^{[l],i_l}$ are tensors and $\lambda^{[l]}$ vectors. If the sites are only slightly entangled, $\lambda^{[l]}_{\alpha_l=1,2\ldots}$ are rapidly decaying numbers and a  cutoff $\chi$ can be introduced in the sum over Greek indices. Moreover, the maximal occupation of the sites can be restricted to $i_l\leq N_{\mathrm{max}}<N$ since it is unlikely that all the bosons occupy a single site. The time evolving block decimation (TEBD) algorithm \cite{Vidal2003,Vidal2004,schollwock11}, which describes how the $\Gamma^{[l]}$ and $\lambda^{[l]}$ evolve in time according to the Bose-Hubbard Hamiltonian, 
is used in order to find the ground state of the system and describe the dynamics. Our tests show that the cutoff values $\chi=500$ and $N_{\mathrm{max}}=9$ 
are sufficient to obtain fully converged results for the temporal dynamics. 
Following Refs.~\cite{carr2001,mishmash2009a,mishmash2009b} we choose as initial state the ground state of the system in the presence of a Gaussian barrier $V(x)=V_0e^{-x^2/\sigma_0^2}$ 
located at the center of the box. For $V_0=18$ and $\sigma_0=0.2$ the one-body density profile is nearly identical to the density profile of the dark soliton described by eq.~(\ref{dsol}). 
At $t=0$ the barrier is turned off and the phase imprinting is applied, with an external potential $V_{\mathrm{ph}}(x,t)=\pi\Theta(x)\delta(t)$, where $\Theta$ is the Heaviside step function and $\delta$ a Dirac pulse. This simply reduces to the transformation 
$\Gamma^{[l],i_l}_{\alpha_{l-1}\alpha_l} \to (-1)^{i_l} \Gamma^{[l],i_l}_{\alpha_{l-1}\alpha_l}$ for the sites with $x>0.$ 

\begin{figure}
\includegraphics[width=0.7\columnwidth]{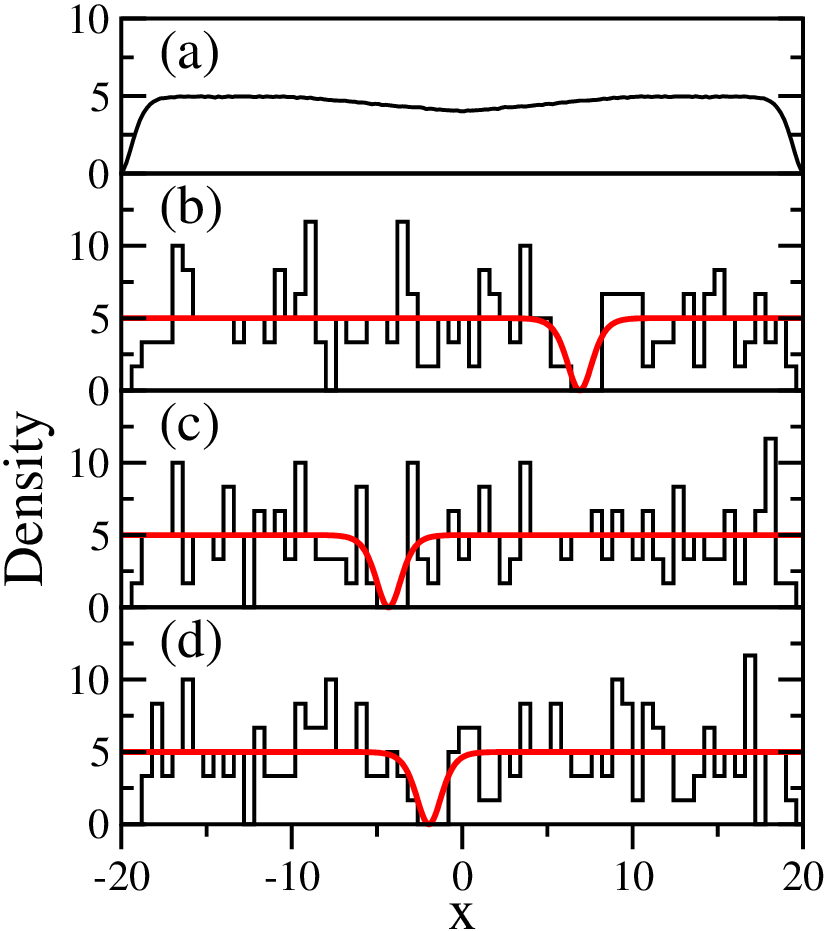}
\caption{(color online) (a) shows the almost uniform one-body density at time $t=25$, while (b)-(d) display the histograms resulting from 3 independent quantum measurements of the positions of all particles.
All the measurements are for the \emph{same} many-body state. Each measurement observes the density notch
associated with the dark soliton at a well defined position, but this position fluctuates randomly for each measurement.
Each histogram can be well fitted by a mean field dark soliton profile (red thick curve), with ``shot noise'' fluctuations
associated with the quantum measurement. Although the one-body density is almost uniform and does not display the existence of the dark soliton, a complete measurement of particle positions reveals its existence.}
\label{outcome}
\end{figure}

In Fig.~\ref{spdensity}, the evolution of the single particle density is presented. Initially, the density follows the prediction of the Bogoliubov approach, with a progressive
filling of the soliton notch. At longer time, the notch is almost completely filled (more than 80\% at $t=25$) and the density tends to be uniform, leading to an apparent
disappearance of the dark soliton. 

In order to gain information on single experimental outcomes, one has to simulate the density measurement in an experiment, 
that is the measurement of the positions of all particles.
In an ideal situation it reduces to the choice of a single Fock state $|i_1,\dots,i_M\ra$ according to the probability density $|\la i_1,\dots,i_M|\psi\ra|^2$. 
This non-trivial task can be performed in a sequential manner in the MPS representation \cite{delande2013}. 

\begin{figure}
\includegraphics[width=0.7\columnwidth]{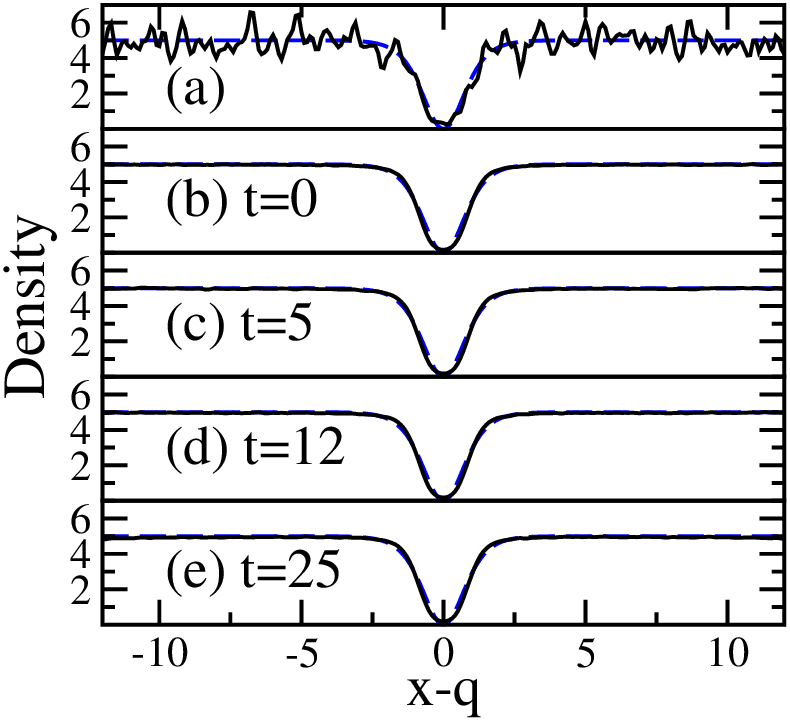}
\caption{(color online) Average density profile with respect to the soliton position $q.$ 
(a) results from the averaging over 50 independent measurements of the same many-body
state at time $t=25.$ Shot noise is still present, but the density notch of the dark soliton clearly emerges.
(b)-(e) are obtained by a massive averaging over $10^5$ independent measurements at various times. 
The shape does not change during the time propagation and is in excellent agreement with the mean field prediction
$|\phi_0(x)|^2,$ eq.~(\ref{dsol}) (blue dashed curve). The dark soliton structure
persists for very long time, even if the soliton position is delocalized over a large range.}
\label{relative}
\end{figure} 

\begin{figure}
\includegraphics[width=0.8\columnwidth]{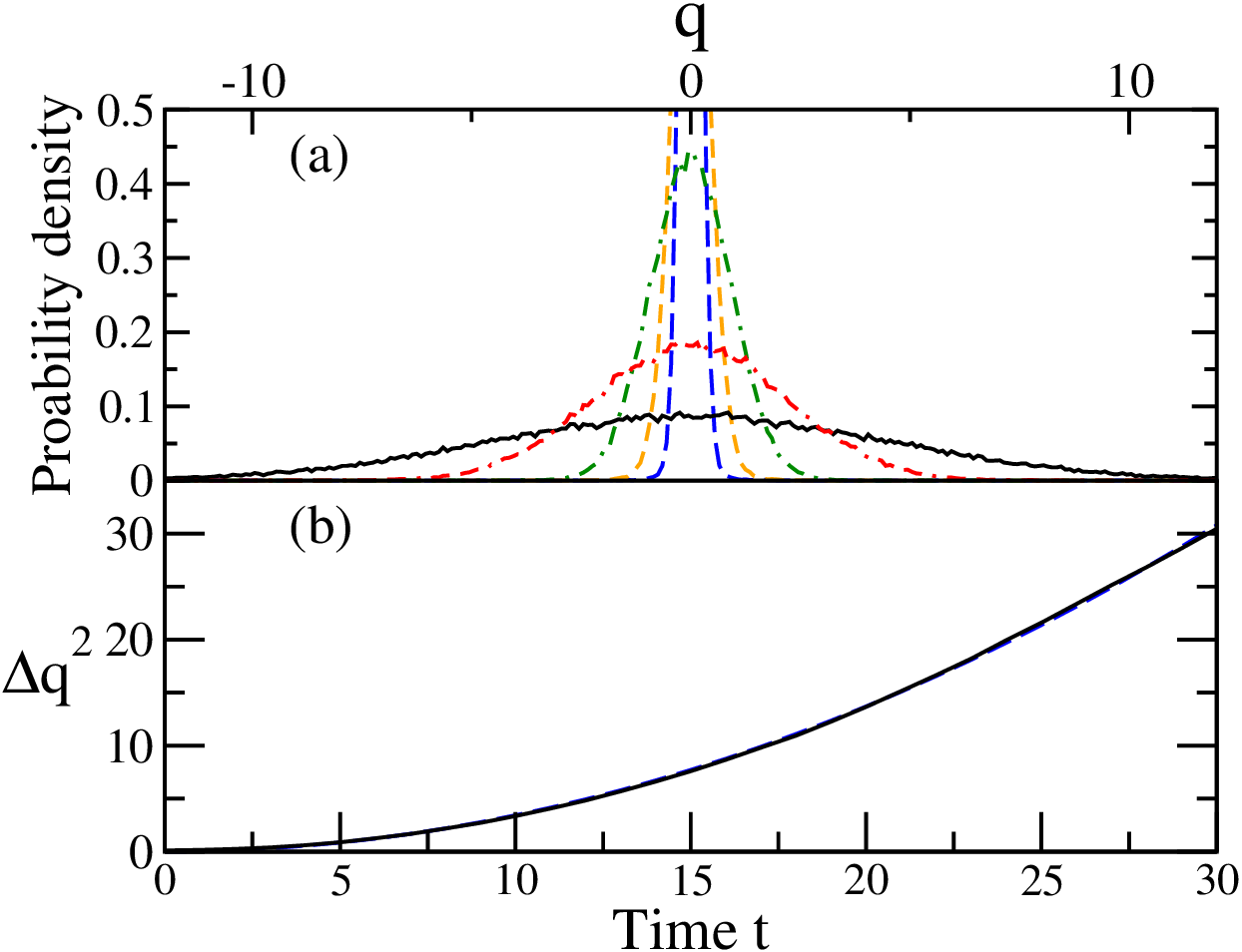}
\caption{(color online) (a): Probability density for the position $q$ of the dark soliton, built
from $10^5$ independent measurements of the same many-body state, at $t=0$ (blue long-dashed), 
2 (orange short-dashed),
5 (green dash-dotted), 12 (red double-dash-dotted) and 25 (black solid). 
Each curve is an almost perfect Gaussian, as predicted
for the spreading of a quantum free wave-packet. This proves that the effective one-body theory, where
the position $q$ of the dark soliton is treated as a quantum degree of freedom, efficiently describes the many-body evolution.
(b): spreading $\Delta q^2$ of the dark soliton position vs. time. For $t>5$, the soliton
is delocalized over a length larger than its size. The prediction of the effective one-body theory, shown
by a blue dashed curve, is almost indistinguishable from the many-body calculation. }
\label{sol_pos}
\end{figure}

In Fig.~\ref{outcome}, we present 3 examples of a single experimental outcome corresponding to $t=25,$ where the
one-body density is almost flat. 
The fluctuations on the number of particles on each site are important: the average number of particles
on each site is 1 for our parameters which makes the detection of the soliton position a bit challenging \footnote{The lattice filling factor is $\rho \Delta x =1,$ similar to the values used in \cite{mishmash2009a,mishmash2009b}. If we could afford a large filling factor as in the case of a model wave-function in \cite{dziarmaga2010}, the dark soliton shape would be clearly visible in a single experiment. However, the value of the filling factor is not a critical parameter in order to observe a dark soliton, cf. \cite{mishmash2010}: the important parameter is the total number of particles in the density notch of the soliton $N_{\mathrm{notch}}=2\rho \xi$. As soon as this number is larger than unity (it is 10 in our case), the soliton position can be identified in particle density measurements.}.
By summing over 3 consecutive sites, i.e. a length $0.6$ (cf. the full soliton size $\approx 2$), ``shot noise'' is reduced, making the soliton position more visible. Clearly, each individual outcome displays a dark soliton 
located at a different random position. These positions can be quantitatively determined by fitting the soliton profile $|\phi_0(x-q)|^2$ to the data, where $q$ is the fitted parameter. Once $q$ is determined, we can shift the density so that the new position of the soliton corresponds to $x=0$. If many results of density simulations, related to the same many-body state, are prepared in this way, we are able to obtain the average profile around the minimum. 
Such a profile, corresponding to 50 realizations is shown in Fig.~\ref{relative}a. It very clearly shows the emergence of a density notch around the soliton position.
When a much larger number of independent realizations is used, the statistical noise becomes negligible, 
and the density profile almost perfectly matches the mean-field density $|\phi_0(x)|^2.$
It is \emph{crucial} to note that the density profiles are the same at different $t$, see
Fig.~\ref{relative}b-e. This means that, even at long time $t=25$ where the single particle density
is almost uniform, the underlying dark soliton is still present and its shape is unaffected,
although its position $q$ fluctuates in different realizations of the measurement. This is a clear signature of the quantum behaviour of the soliton position. 
The simulation of the density measurements involves particle correlations of a very high order 
and cannot be predicted on the basis of the second order correlation function \cite{dziarmaga2010,mishmash2010}.

The slight differences between the average soliton shape and $|\phi_0(x)|^2$, hardly visible in  Fig.~\ref{relative}, are due to the finite particle number. 
The squared overlap between two perfect condensates with slightly different soliton positions, $\Delta q=|q_1-q_2|$ drops like $\exp[-2N_{\mathrm{notch}}(\Delta q)^2/(3\xi^2)]$~\cite{dziarmaga2004}. Thus, if $\Delta q<\xi/\sqrt{N_{\mathrm{notch}}}$, the condensates are not distinguishable and in particular have the same overlaps with any Fock state $|i_1,\dots,i_M\ra$. It means also that the uncertaintity of the determination of the soliton position is of the order of $\xi/\sqrt{N_{\mathrm{notch}}}$. For our parameters the uncertaintity $\Delta q\approx0.3$ and is responsible for the deviation of the average soliton profile 
from the mean field shape.

Simulation of the density measurement also allows us to obtain the probability density of the soliton position, i.e.
how often the soliton localizes at a given position. These densities are shown in Fig.~\ref{sol_pos}. 
In the course of evolution, the histograms become wider and are very well approximated by Gaussian distributions. The variance of the distributions grows like $t^2$ \cite{corney97,corney01}, see Fig.~\ref{sol_pos}b. Thus, the evolution of the probability density resembles the spreading of a free particle Gaussian wave-packet. 

In Ref.~\cite{dziarmaga2004}, an effective one-body (EOB) description of a quantum dark soliton is derived. 
The position and momentum of the soliton are represented by Hermitian operators $\hat q$ and $\hat p_q$, respectively, and its energy by the Hamiltonian $\hat H_q=\hat p_q^2/(2m_{\mathrm{sol}})$, where the dark soliton mass $m_{\mathrm{sol}}=-2N_{\mathrm{notch}}$ is negative. If we want to compare the EOB description with the many-body simulation, we have to determine the initial state $\Psi(q)$ of the quantum soliton. 
This is not so easy if the phase imprinting method is used to excite the system to a dark soliton state. An eigenstate of the $\hat q$ operator is not a satisfactory initial state because it leads to infinite fluctuations of the soliton momentum and consequently to diverging spreading at time $t>0.$ 
However, the many-body calculation shows that the probability density is Gaussian at all times (including at $t=0$),
which shows that the initial state is well approximated by a Gaussian $\Psi(q)=e^{-q^2/2\sigma^2}/(\pi\sigma)^{1/4}$. 
From the previous discussion on the minimum uncertainty of the soliton position, $\sigma$ must be of the
order of $1/\sqrt{N_{\mathrm{notch}}}\approx 0.3.$ The choice $\sigma=0.19$ reproduces
extremely well the observed many-body results, see Fig.~\ref{sol_pos}.
Indeed, within the EOB description, we obtain $\la\hat q(t)^2\ra\approx t^2/(2m_{\mathrm{sol}}^2\sigma^2)$. Considering that the soliton preparation is not perfect (see e.g. radiation of density waves visible in Fig.~\ref{spdensity}) such a quantitative agreement with the EOB theory is remarkable.

In summary, we have considered a dark soliton state in an interacting Bose system in 1D and at zero temperature. In the regime where many-body effects are
important, we show -- within quasi-exact many-body numerical simulations -- 
that the soliton emerges in experiments even though the single particle density shows an almost uniform profile,
in contrast with the claim in~\cite{mishmash2009a,mishmash2009b}. 
The soliton localizes at random positions in different experimental realizations with the 
probability distribution which can be predicted by the effective one-body approach 
where the soliton is considered as a quantum {\it particle} \cite{dziarmaga2004,mochol2012}. 
This proves that, at least on the time scale considered in this letter -- about 10 times longer than the ''quantum soliton lifetime'' claimed in~\cite{mishmash2009a}, Fig. 2 -- the many-body state is a \emph{coherent} superposition of the soliton located at various positions. On a very long time scale, it is not clear how the soliton will lose its phase coherence. The Gross-Pitaevskii equation in 1D is integrable and the soliton cannot decay by emitting low energy excitations \cite{muryshev2002}. Thus, phase coherence could be lost only by truly many-body effects, such as the ones which impose a finite coherence length for a 1D BEC at zero temperature. Investigation of the long time soliton dynamics within the quasi-exact many-body simulations would require computer ressources unfortunately beyond presently available ones.

Another interesting problem, beyond the scope of this paper, is the interaction between two solitons and whether it will be elastic or not. Collision of fast solitons can produce entanglment but it is believed to be elastic \cite{malomedNJP}. On the other hand in \cite{mishmash2009a,mishmash2009b}, a quantum induced inelasticity in soliton collision is claimed, based on pure density plots. In view of the results of the present paper, the density plots
are not conclusive plots and this phenomenon requires further investigation.

We are grateful to Jakub Zakrzewski for discussion. Computing resources have been provided by GENCI and IFRAF. This work was performed within Polish-French bilateral programme POLONIUM No.27742UE. Support of Polish National Science Center via project number DEC-2012/04/A/ST2/00088 (KS) is acknowledged.

\end{document}